\documentclass{article}

    \PassOptionsToPackage{numbers, compress}{natbib}

 \usepackage[final]{neurips_2021}

\usepackage[utf8]{inputenc} 
\usepackage[T1]{fontenc}    
\usepackage{hyperref}       
\usepackage{url}            
\usepackage{booktabs}       
\usepackage{amsfonts}       
\usepackage{nicefrac}       
\usepackage{microtype}      
\usepackage{xcolor}         

\usepackage{comment}

\title{Personalized musically induced emotions of not-so-popular Colombian music}

\author{%
  Juan Sebastián~Gómez-Cañón \\
  MIRLab - Music Technology Group\\
  Universitat Pompeu Fabra\\
  Barcelona, Spain, 08019 \\
  \texttt{juansebastian.gomez@upf.edu} \\
   \And
   Perfecto Herrera \\
   MIRLab - Music Technology Group \\
   Universitat Pompeu Fabra \\
   Barcelona, Spain, 08019 \\
   \texttt{perfecto.herrera@upf.edu}\\
   \AND
   Estefanía Cano \\
   Songquito UG \\
   Erlangen, Germany, 91054 \\
   \texttt{estefania.cano@songquito.com}\\
   \And
   Emilia Gómez \\
   Joint Research Centre\\
   European Commission \\
   Seville, Spain, 41092 \\
   \texttt{emilia.gomez-gutierrez@ec.europa.eu}\\
}

\begin{document}

\maketitle

\begin{abstract}
  This work presents an initial proof of concept of how Music Emotion Recognition (MER) systems could be intentionally biased with respect to annotations of musically-induced emotions in a political context. In specific, we analyze traditional Colombian music containing politically-charged lyrics of two types: (1) vallenatos and social songs from the “left-wing” guerrilla Fuerzas Armadas Revolucionarias de Colombia (FARC) and (2) corridos from the “right-wing” paramilitaries Autodefensas Unidas de Colombia (AUC). We train personalized machine learning models to predict induced emotions for three users with diverse political views -- we aim at identifying the songs that may induce negative emotions for a particular user, such as anger and fear. To this extent, a user's emotion judgements could be interpreted as problematizing data -- subjective emotional judgments could in turn be used to influence the user in a human-centered machine learning environment. In short, highly desired “emotion regulation” applications could potentially deviate to “emotion manipulation” -- the recent discredit of emotion recognition technologies might transcend ethical issues of diversity and inclusion.
\end{abstract}

\section{Introduction}
From the computational perspective,  Music Emotion Recognition (MER) attempts to predict the emotion perceived by or induced in a particular listener \citep{Yang2011}. 
Despite the criticism to MER due to subjectivity (and response diversity) \citep{GomezCanon2021SPM}, a growing effort has been made to produce enriched datasets of emotion judgements with more listening data to better represent the properties and context of the listener \citep{Barthet2013,Schedl2013}: demographics, cultural and individual differences, preference, familiarity, functional uses of music, physiological signals, and language. 
In this context, personalized and context-sensitive models that incorporate this information could more accurately predict the particular emotion judgments from a particular listener or groups of listeners \citep{Yang2007,GomezCanon2020ISMIR,GomezCanon2021ISMIR}. 
However, in the context of evolving technologies for user profiling, the High Level Expert Group of the European Commission proposed that societal well-being is one of seven requirements to produce trustworthy artificial intelligence systems\footnote{\url{https://op.europa.eu/s/pInE}} -- MER systems could potentially be harmful when inducing particular emotions to a listener \citep{GomezCanon2021SPM}. 

The aim of this paper is to attempt to understand if and under which circumstances MER algorithm can effectively be biased to recommend music that could induce negative emotions on a particular listener. 
Namely, a personalized MER algorithm that produces music recommendations grounded on the users' emotional judgments (an already standardized practice in music streaming platforms), could influence the user either positively or negatively. 
This study diverges from previous research which aims at using music to enhance memory, relieve boredom, improve concentration, promote prosociality, or aid learning \citep{Hu2021}. 
The reason is that, despite the common consensus regarding beneficial uses of music, generalized misconceptions have surged historically in the perception of well-being applications of music (e.g., the Mozart effect \citep{Mehr2013}, binaural beats \citep{Orozco2020}, or 432 Hz music \citep{Rosenberg2021}). 
Only recent research has started to theorize and analyze music-induced harm \citep{Ziv2016,Silverman2020} -- a topic that should be more widely discussed and studied by academia given the possibility that social networks, streaming platforms, and personalized advertisement companies are already studying it and potentially gaining profit from \citep{ONeil2016,Noble2018,Zuboff2019,Veliz2020}. 

\section{Related work}

\subsection{Music and Emotions}

Despite longstanding debates around the emotions that music can evoke, the appealing nature of human emotion studies has sparked what has been recently referred to as “the rise of affectivism” \citep{Dukes2021}: methodological and technical advancements of the emotion studies have surged to gain deeper understanding of behavioral and cognitive processes. 
With respect to music, \textit{induced} emotions concern the arousal of psycho-physiological responses to a particular stimuli \cite{Krumhansl1997}. 
We refer the reader to \citep{Meyer1956,Kivy1990,Juslin2010Handbook,Juslin2019,Eerola2013,eerola2018handbook,Warrenburg2020} for extended theory and research on this  topic.
We use a discretized model of emotion based on Russell's circumplex model \citep{Russell1980} and recent work on MER \citep{Panda2018}, which conceptualizes emotions in two dimensions (i.e., arousal and valence) and four distinct categories/quadrants of emotion: $Q_1$ (positive valence and arousal), $Q_2$ (positive arousal and negative valence), $Q_3$ (negative valence and arousal), $Q_4$ (negative arousal and positive valence). 
Arousal refers to energy or activation and valence relates to pleasantness or positiveness of an emotion: $Q_1$ refers to emotions such as joy, wonder, and power; $Q_2$ refers to emotions as tension, anger, and fear; $Q_3$ refers to emotions as sadness, bitterness; $Q_4$ refers to emotions as tenderness, peacefulness, and transcendence \citep{Zentner2008}.  
For this particular study, we focus on music that can induce $Q_2$-emotions for a particular user. 

\subsection{Use case: Polarization in Colombian music}

The contextualization of the political landscape in Colombia escapes the scope of this paper, thus we refer the reader to \citep{chomsky2004,Zamosc1986,Berquist1978,Stokes2005,Arocha1988,FalsBorda2001,Mahoney2020} for deeper analysis regarding the history of violence in Colombia. 
As context, the “biblical holocaust” of Colombian violence -- portrayed by the writer Gabriel García Márquez -- has resulted in more than 420,000 violent deaths over the last 70 years, more than 11 million Colombians leaving the country or internally displaced, and one of the most unequal distribution of income in the continent \citep{Mahoney2020}. 
Diverse sources of inequalities (e.g., agrarian capitalism, socioeconomic exclusion, decolonization processes, the war on drugs, illegal economies, and exploitation of natural resources) are the cause of the formation of illegal armies fueled by political ideologies \citep{Grajales2021}: “left-wing” Fuerzas Armadas Revolucionarias de Colombia (FARC) and “right-wing” Autodefensas Unidas de Colombia (AUC), amongst several other illegal groups. 
As an oversimplification of Colombia's historical process (and reflecting the generalized trend in the world), polarization arose over whether and how to pursue peace in the country, producing negative relationships between political discourses and everyday life \citep{Feldmann2019}.
Polarization results in fragile societies, cooperation within citizens becomes complicated, and collective action for problem solving becomes impossible \citep{Veliz2020}. 
We argue that the music selection, described as follows, is \textit{not-so-popular} Colombian music since the political content of the lyrics may arouse arguments from different political views. 

\section{Methodology}

\subsection{Music selection}
 
We refer the reader to studies by \citet{Quishpe2020}, \citet{BarbosaCaro2019}, and \citet{KatzRosene2017} with respect to historical, functional, and lyrical analysis from the two types of music used: (1) FARC-songs (mainly in the style of \textit{vallenato} and \textit{canción social}) and (2) AUC-songs (in the style of \textit{corridos}). 
These musical styles make part of traditional Colombian (and Latin-american) music, yet they have distinctive sonorities, structures, and instrumentation. 
It must be noted that music with politically motivated lyrics from both types have used similar styles of music as well (e.g., hip-hop and rock), but this study only considers this reduced range of styles. 
Additionally, FARC-songs have been typically created by active members from the guerrilla as a mechanism of identity confirmation and propaganda \citep{Quishpe2020}, while AUC-songs have been typically produced by sympathizers of the paramilitaries as promotion to their deeds and open crtiticism to the FARC \citep{BarbosaCaro2019} -- the functionality of the music and the target listener are different. 

We remark that humans frequently listen to music \textit{without} feeling any emotion at all \citep{Kivy1990,Juslin2019}, but we assume that music \textit{might} trigger episodic memories to particular individuals \citep{Juslin2013,eerola2018handbook}. 
The potential induction of emotions from this music is due mainly to the semantic content of the lyrics (see \ref{sec:future_work}) -- inducing different emotions to listeners with different political views. 
Despite the importance of lyrics to the induced emotions, we argue that the acoustic features are useful to provide a content-based contrast among the different styles of music: (1) FARC-songs typically use less instruments and might include only voice and guitar, and (2) AUC-songs are more heavily orchestrated with faster tempo. 
We use 50 music excerpts with lyrics from each music type (30 seconds long) and extracted 260 emotionally relevant acoustic features (mean and standard deviation of 65 low-level music descriptors and their first order derivatives) from segments of 1 second \cite{Aljanaki2017}, with 50\% overlap, and standardize across features -- using the IS13 ComParE feature set \cite{Weninger2013} and OpenSMILE toolbox \cite{openSMILE2013}. 
Namely, the machine learning models should be able to differentiate between the types of music -- the interesting element is to attempt to understand which users will provide problematizing labels (i.e., music that induces subjective emotions of anger or fear) that can bias the algorithm towards a particular class. 

\subsection{Classification strategy}

We use the “machine consensus” MER personalization strategy presented by \citet{GomezCanon2021ISMIR}: consensus entropy for active learning. 
This strategy uses a committee of classifiers to analyze their output agreement and queries each user for instances with the highest uncertainty. 
A committee of classifiers (15 independent Extreme Gradient Boosting models \cite{Chen2016xgb}) has previously been pre-trained on separate cross-validation splits of the DEAM dataset, the benchmark dataset for MER \citep{Aljanaki2017}. 
In order to select uncertain data to be labeled, classifiers predict the output probabilities for the pool of excerpts. 
We then perform the consensus entropy strategy by analyzing the disagreement across classifiers. 
For example, full disagreement from a committee of four classifiers results when each one predicts a different class/quadrant with 100\% probability. 
This yields average probabilities per quadrant $p_{avg} = \{Q_1: 0.25, Q_2: 0.25, Q_3: 0.25, Q_4: 0.25\}$ and high inter-class entropy/uncertainty of 1.386. 
Excerpts with highest uncertainty are then queried to the oracle (i.e., each user) to be annotated.
Initially, we randomly draw 5 excerpts from each type of music (10 excerpts for the first annotation iteration), retrain our classifiers with the annotations provided by each user, identify the excerpts to be annotated for the next iteration, and present the new batch of music to be annotated. 
Given the low amount of available music, we perform only three iterations for a total of 30 annotations per user -- past research has shown that only 20-30 annotations are needed in order to reach personalization \citep{Su2012,Chen2017}. 
Please refer to \citep{GomezCanon2021ISMIR} for additional information of the methodology. 

\section{Preliminary results and Discussion}

We test our initial models with three users who have reported their political opinions: one with “left-wing”, one with “right-wing”, and one with “center” political views \citep{Feldmann2019}. 
We ask our users to report their possible vote in a forced choice question -- if the elections would happen now for whom would they vote? 
They were able to choose between the “left-wing” candidate from the Colombia Humana party (Gustavo Petro), a “right-wing” candidate from the Centro Democrático government party (there is no clear candidate yet), or cast a blank vote. 

\subsection{$Q_2$ music prediction}
Given that this study is not centered on improving the accuracy of models but rather understanding how the models are progressively biased towards a particular category, we test the finalized models on the music which has not been annotated by the user. 
Namely, each user annotates 30 excerpts (3 iterations of 10 annotations each) and we test the personalized model on the remaining testing data (70 excerpts), which are different for each user. 
In particular, our interest is to study if a personalized model trained for a user with a particular political view, can effectively identify new music that can potentially induce negative emotions (i.e., music belonging to $Q_2$). 
We obtain output probabilities for the testing data, sort the highest probabilities for $Q_2$, and count the amount of songs that belong to each political view from the top 10 predictions. 
Additionally, we report the mean output probability $p_{avg}$ of each type of music belonging to $Q_2$ (see Table \ref{tab:results}). 

\begin{table}[ht!]
  \caption{Preliminary results of personalized models.}
  \label{tab:results}
  \centering
  \begin{tabular}{lllll}
    \toprule
      & \multicolumn{2}{c}{Top 20} & \multicolumn{2}{c}{Output $p_{avg}$}    \\
        \cmidrule(r){2-3}
        \cmidrule(r){4-5}
    User type & FARC-songs & AUC-songs & FARC-songs & AUC-songs \\
    \midrule
    “Left-wing”  & 10\% & 90\% & 93.37\% & 71.02\% \\
    “Center”     & 30\% & 70\% & 90.57\% & 34.49\% \\
    “Right-wing” & 70\% & 30\% & 92.60\% & 72.76\% \\
    \bottomrule
  \end{tabular}
\end{table}

The initial remarks are that, for this reduced number of users, the personalized models for users with distinctive political views (i.e., “right-wing” and “left-wing”) appear to capture that music from the opposite political perspective will have more likelihood to produce $Q_2$ emotions. 
While the amount of participants is not indicative of generalizable trends, we argue that these personalized models could indeed be “breached” with problematizing subjective data -- testing data was completely unseen by the users, relying exclusively on acoustic features which are not the reason for inducing emotions. 
In fact, further analysis with $p_{avg}$ reveals that music of both types is very likely to produce $Q_2$-emotions -- in fact, the assumption that the music selection is \textit{not-so-popular} might hold.
As a limitation, we identify that this methodology can be randomly determined by the selection of songs for the initial iteration (see \ref{sec:future_work}). 

\subsection{Future work and broader impact}\label{sec:future_work}

Our final interest is to understand whether the use of this music can potentially have an impact on decision-making -- studies regarding the impact of musically-induced emotions on decision making have been studied since 2006, as reviewed by \citet{Palazzi2019}. 
It has been argued that music and persuasion have indirect relationships and are never strictly causal -- as the persuasion/manipulation of a person is already a difficult task, it can only be  “helped” or “promoted” by music. 
As mentioned by \citet{Herrera2009}, music can only contribute as a “persuading factor” to an induced emotional state, which can be associated between music and a particular person or message, contributing to re-evaluating attitudes and actions. 
In fact, the author stresses the fact that music containing a message within the lyrics can “make the complete brain work in a coordinated manner”: text and lyrics will activate more the left hemisphere, while music will activate the right one. 
Given that the main emotion-inducing mechanism is the semantic content of the lyrics for this study, we plan to add multi-modality to this approach by using sentiment analysis models on the lyrics. 
In principle, sentiment analysis models could also be re-trained with new annotations that could bias the ensemble model to identify this problematic data. 
Furthermore, the variability of the initial annotations for personalization and the impact of the initial iteration on concept drift \citep{Widmer1996}, must still be evaluated and analyzed.  

In summary, it is very likely that any type of stimuli that produces strong political responses can be somewhat captured by the personal annotations used as input to a machine learning model. 
It has been argued that music is a powerful and engaging stimulus that can promote prosociality \citep{Ruth2018}, impact customer behaviors \citep{Hansen2007}, and influence processes of decision-making and risk-aversion \citep{Fischer2006,Greitemeyer2011}. 
We want to understand how the use of growing emotion recognition technologies can have a direct impact on ethical issues, mainly societal well-being -- human-centered technologies must be evaluated for both beneficial and harmful use cases. 

\begin{ack}
The research work conducted at the Universitat Pompeu Fabra is partially supported by the European Commission under the TROMPA project (H2020 770376) and the Project Musical AI - PID2019-111403GB-I00/AEI/10.13039/501100011033 funded by the Spanish Ministerio de Ciencia, Innovación y Universidades (MCIU) and the Agencia Estatal de Investigación (AEI).
\end{ack}


\medskip

\small

\bibliography{references}

\end{document}